\documentclass{emulateapj}
\usepackage{apjfonts}

\def\etal{{\it et~al.}}

\shorttitle{A Dense Gas Trigger for OH Megamasers}
\shortauthors{Darling}

\begin{document}
\title{A Dense Gas Trigger for OH Megamasers}
\author{Jeremy Darling\altaffilmark{1}}
\altaffiltext{1}{Center for Astrophysics and Space Astronomy,
Department of Astrophysical and Planetary Sciences,
University of Colorado, 389 UCB, Boulder, CO 80309-0389; 
jdarling@origins.colorado.edu}

\begin{abstract}
HCN and CO line diagnostics provide 
new insight into the OH megamaser (OHM) phenomenon, suggesting a 
dense gas trigger for OHMs.  
We identify three physical properties that differentiate
OHM hosts from
other starburst galaxies:  
(1) OHMs have the highest mean molecular gas densities among starburst
galaxies; nearly all OHM hosts have
$\bar{n}({\rm H}_2)=10^3$--$10^4$~cm$^{-3}$ (OH line-emitting clouds likely 
have $n({\rm H}_2)>10^4$~cm$^{-3}$).
(2) OHM hosts are a distinct population in the nonlinear part of the 
IR-CO relation.
(3) OHM hosts have exceptionally high dense molecular gas fractions,
$L_{\rm HCN}/L_{\rm CO}>0.07$, and comprise roughly half of this 
unusual population.
OH absorbers and kilomasers generally follow the linear IR-CO
relation and are uniformly distributed in dense gas fraction and
$L_{\rm HCN}$, demonstrating that OHMs are independent of OH abundance.
The fraction of non-OHMs with high mean densities
and high dense gas fractions constrains beaming to be a minor effect:
OHM emission solid angle must exceed $2\pi$ steradians.
Contrary to conventional wisdom, 
IR luminosity does not dictate OHM formation; both star formation and 
OHM activity are consequences of tidal density enhancements accompanying 
galaxy interactions.  The OHM fraction in starbursts is likely
due to the fraction
of mergers experiencing a temporal spike in tidally driven density  
enhancement.  
OHMs are thus signposts marking the most intense, compact, and 
unusual modes of star formation in the local universe.
Future high redshift OHM surveys can now be interpreted in a star 
formation and galaxy evolution context, indicating both the merging rate
of galaxies and the burst contribution to star formation.
\end{abstract}

\section{Introduction}

OH megamasers (OHMs) are rare luminous 18~cm masers associated with 
major galaxy merger-induced starbursts.
The hosts of OHMs are (ultra)luminous IR galaxies ([U]LIRGs), and the 
OHM fraction in (U)LIRGs peaks at about 1/3 
in the highest luminosity mergers \citep{darling02a}.  
It is not known whether all major mergers experience an
OHM stage or what detailed physical conditions produce OHMs, but
it is clear that OHMs are a radically different phenomenon from 
the aggregate OH maser emission associated with ``normal'' (Galactic) 
modes of star formation in galaxies.
\citet{lo05} posed a key question:
why do $80\%$ of LIRGs show no OHM 
activity?  To reframe the question:  given two merging systems with 
similar global IR and radio continuum properties in the same morphological
stage of merging, why does one show OHM emission while the other does not?
What is the difference between the two systems?  Perhaps there is no 
difference and the fraction of OHMs among mergers simply reflects beaming
or OH abundance.
Or perhaps OHM activity depends on small-scale conditions that are 
decoupled from global properties of mergers.

The provenance of OHM emission vis-\`{a}-vis the host galaxy 
has been extensively investigated in the radio through
X-ray bands by comparing samples of OHM galaxies to 
similarly selected non-masing control samples.
For example, 
\citet{darling02a} and \citet{baan06} studied 
radio and IR properties vis-\`{a}-vis the AGN versus starburst contributions
to OHM activity, 
\citet{baan1992} and \citet{darling02a}
investigated the OHM fraction in (U)LIRGs versus star
formation rate and IR color, 
\citet{baan1998} and \citet{darling06} used optical spectral 
classification to distinguish populations and to quantify AGN fraction in 
OHM hosts, 
and \citet{vignali05} conducted an X-ray study of
the contribution of AGNs to OHM hosts.
While some of these studies pointed to minor
differences in statistical samples of OHM hosts versus nonmasing systems, 
they could not identify on a case-by-case basis which systems would 
harbor OHMs and which would not based on any observable quantity except the 
OH line itself.  

Theoretical modeling of OHM formation has seen a recent renaissance:
\citet{parra2005} model the $\sim50$~pc molecular torus in 
III~Zw~35 and show how 
OHM emission is a stochastic amplification of unsaturated emission by 
multiple overlapping clouds, and \citet{lockett07}
show how the general excitation of OHMs is fundamentally
different from Galactic OH maser emission and predict that a single 
excitation temperature governs all 18~cm OH lines.  While the physics of OHMs
is crystallizing, and models predict that beaming is not likely to be
the dominant factor in the OHM fraction among (U)LIRGs, it remains unclear 
on a case-by-case basis what conditions found in starbursts drive or prohibit
OHM formation.  


Here we describe a dense gas trigger for OHM formation, at last
identifying physical observable properties that differentiate
OHMs from nonmasing mergers.  We identify OHMs, 
OH absorbers, OH kilomasers, and OH non-detections in the 
Gao \& Solomon (2004a; hereafter GS04a) sample (\S \ref{sec:sample}) and 
employ CO($1-0$) and HCN($1-0$) molecular gas tracers 
to show that while OH absorbers appear nearly uniformly distributed 
in $L_{\rm IR}$ and $L_{\rm HCN}$, OHMs represent the {\it majority} of
the nonlinear population in the IR-CO 
relation (\S \ref{sec:results}).  In combination with 
a Kennicutt-Schmidt-based star formation model of CO line emission
by \citet{krumholz07}, we identify a high mean molecular density 
driving OHM emission, and from the HCN/CO ratio
we find that OHM galaxies are exclusively 
high dense gas fraction starbursts (\S \ref{sec:results}).
Now that we can at last observe quantities
that are highly predictive of OHM activity, we can employ OHMs at 
high redshifts as probes of major galaxy mergers and extreme 
star formation (\S \ref{sec:discussion}).

\section{The Sample}\label{sec:sample}

\begin{figure}
\epsscale{1.2}
\plotone{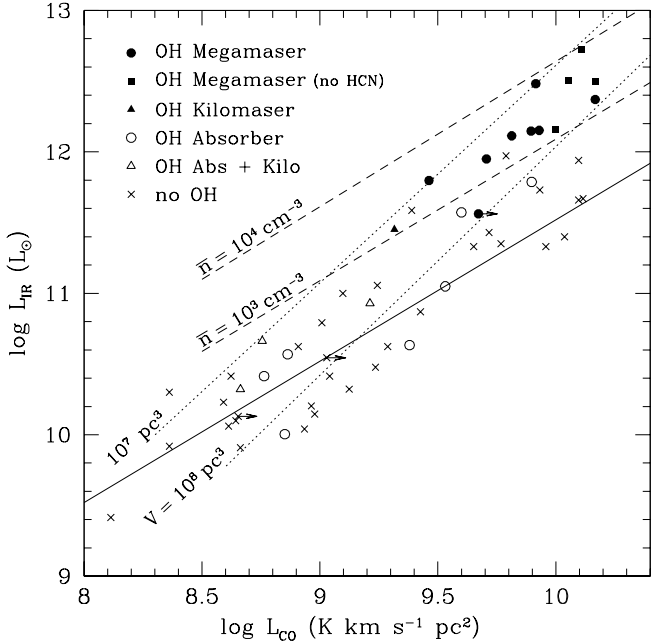}
\caption{
IR luminosity versus CO line luminosity in 
HCN-detected galaxies with known OH properties 
from the GS04a sample.
The legend indicates symbols for OH megamasers, 
OH kilomasers, OH absorbers, and objects with no
detected OH lines.   The solid line is a linear 
fit by \citet{gao2004b} to galaxies with $L_{\rm IR}<10^{11} L_\odot$
($L_{\rm IR}= 33 L_{\rm CO}$ in units above), the dotted lines indicate 
a constant total volume of molecular material, and the dashed
lines indicate the mean H$_2$ density derived from \citet{krumholz07}.
\label{fig:IR_vs_CO}}
\end{figure}

The somewhat diverse GS04a HCN sample that forms the basis for this study 
includes most IR- and CO-bright 
galaxies (by flux) and most local northern ULIRGs ($cz < 20,000$~km~s$^{-1}$).
Table \ref{tab:HCNsample} lists basic properties and line luminosities
of all galaxies in the sample
that have been observed in the 1667~MHz OH line by various groups.
The HCN sample includes 8 OHMs, 
12 OH absorption systems, 4 OH kilomasers, and 40 OH nondetections. 
While the division between OH kilomasers and OHMs 
at $L_{\rm OH} = 10~L_\odot$ is rather arbitrary, 
the $L_{\rm OH}$ values 
for the OH kilomasers are well separated from the OHMs in Table 
\ref{tab:HCNsample} by 2 orders of magnitude.
Three of the four OH kilomasers in this sample show both emission and 
absorption.  OH types marked with an asterisk are somewhat uncertain
and have been omitted from all subsequent analysis and figures.
We have included in Figure \ref{fig:IR_vs_CO} four additional OHMs
that have been detected in CO by \citet{solomon97} but have not yet been 
observed in HCN:  IRAS~03521+0028, 14070+0525, 16090$-$0139, and 18368+3549.




\begin{figure}
\epsscale{1.2}
\plotone{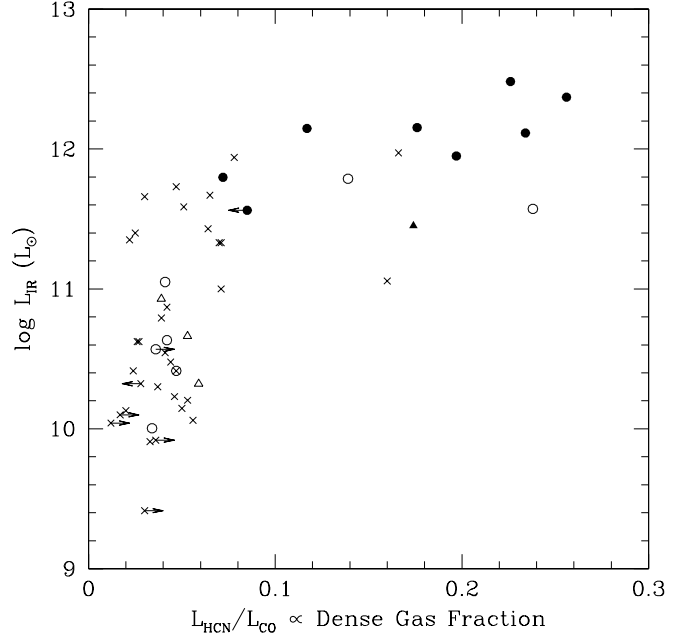}
\caption{
IR luminosity versus $L_{\rm HCN}/L_{\rm CO}$,
a proxy for the dense gas fraction, 
in HCN-detected galaxies with known OH properties 
from the GS04a sample.
Symbols are identical to those used in Figure \ref{fig:IR_vs_CO}.
\label{fig:IR_vs_dense}}
\end{figure}

\section{Results}\label{sec:results}


Sorting the GS04a sample by OH type --- megamaser, kilomaser, 
absorber, or non-detection --- 
reveals striking properties of OHM host galaxies that set them apart from
other starburst galaxies.  
Figure \ref{fig:IR_vs_CO} shows that OHMs comprise the majority 
of the population that is offset from the 
linear $L_{\rm IR}$-$L_{\rm CO}$ relation.
OH absorbers and kilomasers, however, generally follow the linear IR-CO 
relation.
Using the relationship between $L_{\rm IR}/L_{\rm CO}$ and the mean
H$_2$ density, $\bar{n}$, derived from
Kennicutt-Schmidt laws by \citet[][Fig.\,2]{krumholz07}, we show in Figure 
\ref{fig:IR_vs_CO} that all
OHMs in the sample are produced in starburst volumes of $10^7$--$10^8$~pc$^3$
(radii $\sim$130--290~pc) and that all but one OHM have extremely
high volume-averaged molecular densities, $\bar{n}=10^3$--$10^4$~cm$^{-3}$.
In fact, 7 of 10 objects in the HCN sample in this density range are OHMs and
one is a nonabsorbing OH kilomaser.  Note that $\bar{n}$ is the {\it mean}
H$_2$ density; the clouds responsible for OHMs within these
regions must be significantly denser than the mean.  
There are many nonmasing systems with high $L_{\rm IR}$ at
lower densities and larger volumes than the OHMs, 
demonstrating that molecular density, not the IR radiation field, is the OHM
trigger.

\begin{figure*}
\plotone{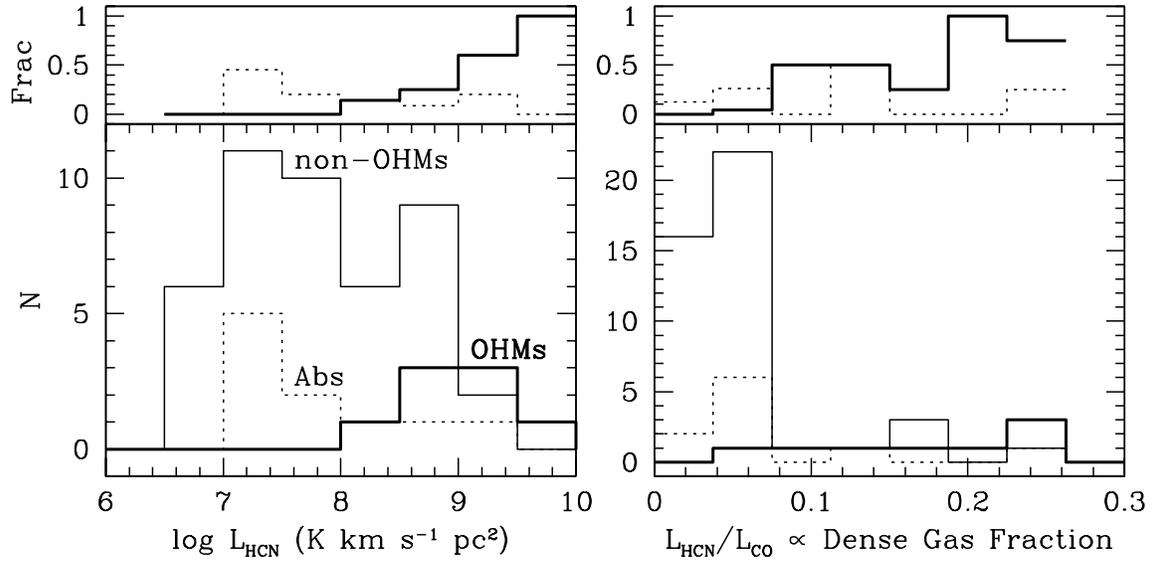}
\caption{
Number and fraction of OH megamasers, OH absorbers, and non-OHMs
(including absorbers) versus $L_{\rm HCN}$ and $L_{\rm HCN}/L_{\rm CO}$, a
proxy for dense gas fraction.  
The bold line indicates the OHMs, the solid line indicates the non-OHMs, and
the dotted line shows the OH absorbers, as indicated in the left panel.
The upper panels indicate the fraction of the total sample that shows
either OH absorption or emission.
\label{fig:hists}}
\end{figure*}

Equally striking is the segregation of OHMs from nonmasing starbursts
in a plot of $L_{\rm IR}$ versus $L_{\rm HCN}/L_{\rm CO}$, a proxy for 
dense molecular gas fraction \citep{gao2004b}.
Figures \ref{fig:IR_vs_dense} and \ref{fig:hists} show that
all 8 OHMs (and the nonabsorbing OH kilomaser) 
have $L_{\rm HCN}/L_{\rm CO} > 0.07$.  There are 
also 2 OH absorbers and 5 OH nondetections this regime, so 
OHMs comprise roughly half of this unusual population.
There are no OHMs with $L_{\rm HCN}/L_{\rm CO}<0.07$, 
but there are many other luminous systems,
including OH absorption systems and OH absorbers with
coincident kilomaser emission.  Galaxies with OH absorption are fairly 
uniformly distributed in $L_{\rm IR}$ and dense gas fraction,
indicating that OH abundance is not a factor in OHM formation.
The $\gtrsim50\%$ fraction of OHMs in dense starbursts
constrains OHM beaming be a minor contributor
to the OHM fraction in LIRGs, and we conclude that the 
OHM emission solid angle must be greater than $2\pi$ steradians.  







\section{Discussion}\label{sec:discussion}

A dense gas trigger for OHM formation
is consistent with the modeling work by \citet{parra2005} showing 
that the critical component for OHM formation is cloud-cloud overlap;
a starburst-scale high mean molecular density and 
high dense gas fraction both provide the required overlap of 
many dense clouds.  What is not yet clear is whether OHM activity is a density
effect or simply a concentration effect.  The rough size scales
bracketing all OHMs in Figure \ref{fig:IR_vs_CO} are also consistent 
with the 100--200~pc OHM emission regions observed with VLBI
\citep[e.g.,][]{pihlstrom01,rovilos03}.


It is somewhat surprising that significant differences between OHM hosts and 
nonmasing systems are seen at all in unresolved observations
because masing nuclei are ``contaminated'' by nonmasing nuclei within mergers.
We expect that the observed differences between OHMs and nonmasing starbursts
would intensify in resolved observations.  However, 
high dipole moment molecules such as 
HCN and CS may be good OHM location-selective tracers that obviate the need
to obtain subarcsecond resolution.

It is certain that the OHM phase is transitory because star formation rates
found in these systems are sustainable for $10^7$--$10^8$~yr, 
whereas the complete merging process requires of order $10^9$~yr.
What is not known, however, is whether this mode of star formation 
is a universal, inevitable stage or an uncommon event in major galaxy mergers.  
Tidal torques spike multiple times during 
mergers, and many major mergers are likely to experience a ULIRG phase, 
but will most mergers experience the even more extreme OHM phase? 
The simple 
observation that about $20\%$ of LIRGs with $L_{\rm FIR} > 10^{11.2}L_\odot$ 
show OHM activity \citep{darling02a} suggests that
if {\it all} LIRGs experience an OHM stage, then the OHM lifetime is of 
order $20\%$ of the LIRG lifetime.  If only a subset of LIRGs have an OHM
stage, then the OHM lifetime must be longer.  It is also possible that 
OHMs draw from a larger ``pool'' of galaxies that begin at lower $L_{\rm FIR}$
than the LIRG sample, which would allow a shorter OHM lifetime.  
Constraints on the OHM lifetime are clearly critical to understanding 
OHMs and the modes of star formation in major galaxy mergers.



\section{Conclusions}

We have identified three closely related
physical properties that differentiate OHMs 
from other starburst galaxies:  
OHM hosts have the highest mean molecular gas densities, 
they are a distinct population in the nonlinear part of the IR-CO relation, 
and they 
reside in galaxies with exceptionally high dense molecular gas fractions.
We conclude that molecular gas must be concentrated and massive in order to 
reach the mean density required to form an OHM in a galactic nucleus.
IR luminosity is not a condition for OHM formation; both star formation and 
OHM activity are consequences of the tidal density enhancements accompanying 
galaxy interactions.  
The fraction of OHMs in dense starbursts
constrains OHM beaming to be a minor effect:
OHM solid angle emission must be greater than $2\pi$ steradians.  
These conclusions are in good agreement with the stochastic cloud-cloud overlap
amplification model by \citet{parra2005}.
The rather uniform distribution of OH absorbers in IR, HCN, and CO luminosity
suggests that OH abundance is not a significant factor in OHM formation.
The main caveat to these conclusions is that 
the sample of OHMs with HCN observations remains small, and should be 
expanded, particularly to higher redshifts to include ``typical'' OHMs.

OHMs are signposts of the most intense, compact, and 
unusual modes of star formation in the local universe, and 
surveys for OHMs will now provide detailed information
about the detected host galaxies and their mode of star formation.  
The missing datum required for a complete interpretation of 
OHM surveys, however,  is the OHM lifetime.  


\acknowledgements
This work benefited significantly from comments by the anonymous referee.
This research has made use of the NASA/IPAC Extragalactic Database (NED),
which is operated by the Jet Propulsion Laboratory, California Institute of
Technology, under contract with NASA.

\begin{deluxetable}{llccccccc} 
\tabletypesize{\scriptsize}
\tablecaption{OH Properties of HCN-Detected Galaxies\label{tab:HCNsample}}
\tablewidth{0pt}
\tablehead{
\colhead{IRAS Name} &  
\colhead{Other Name} &  
\colhead{$D_L$} & 
\colhead{$L_{\rm IR}$} & 
\colhead{$L_{\rm CO}$} & 
\colhead{$L_{\rm HCN}$} & 
\colhead{OH Type\tablenotemark{a}} & 
\colhead{$\log L_{\rm OH}$} & 
\colhead{Ref} \\
\colhead{} & 
\colhead{} & 
\colhead{(Mpc)} & 
\colhead{($10^{10} L_\odot$)} & 
\colhead{($10^8$ K km s$^{-1}$ pc$^2$)} & 
\colhead{($10^8$ K km s$^{-1}$ pc$^2$)} & 
\colhead{} & 
\colhead{($L_\odot$)} 
}
\startdata
00450$-$2533 & NGC253 &  2.5 &  2.1 &  4.6 & 0.27 & abs+kilo& ($-$1.3) & 1\\
01053$-$1746 & IC1623 & 81.7 & 46.7 &130.5 & 8.5  & non & \nodata & 2\\
01219+0331 & NGC520 & 31.1 &  8.5 & 16.3 & 0.64 & abs+kilo & ($-$0.4) & 3\\
01403+1323 & NGC660 & 14.0 &  3.7 &  7.3 &$>$0.26& abs& \nodata & 4\\
02071+3857 & NGC828 & 75.4 & 22.4 & 58.5 &  1.3 & non & \nodata & 2 \\
02193+4207 & NGC891 & 10.3 &  2.6 & 11.0 & 0.25 & non & \nodata & 2 \\
02360$-$0653 & NGC1022& 21.1 &  2.6 &  4.2 & 0.20 & non & \nodata & 2 \\
02391+0013 & NGC1055& 14.8 &  2.1 & 13.3 &$<$0.37&non & \nodata & 5\\
02401$-$0013 & NGC1068& 16.7 & 28.3 & 20.7 & 3.61 & kilo & ($-$0.3) & 6\\
02526$-$0023 & NGC1144&117.3 & 25.1 &108.9 & 2.67 & non & \nodata & 2 \\
03317$-$3618 & NGC1365& 20.8 & 12.9 & 58.7 & 3.10 & non*& \nodata & 7\\
03419+6756 & IC342  &  3.7 &  1.4 &  9.5 & 0.47 & non & \nodata & 2 \\
04315$-$0840 & NGC1614& 63.2 & 38.6 & 24.5 & 1.25 & non & \nodata & 2 \\
05083+7936 & VIIZw31&223.4 & 87.1 &125.0 & 9.8  & non & \nodata & 2 \\
05189$-$2524 & \nodata&170.3 &118.1 & 67.0 & 6.2  & non*& \nodata & 7\\
06106+7822 & NGC2146& 15.2 & 10.0 & 12.5 & 0.96 & non & \nodata & 2 \\
07101+8550 & NGC2276& 35.5 &  6.2 & 10.2 & 0.40 & non & \nodata & 2 \\
09126+4432 & Arp55  &162.7 & 45.7 &125.0 & 3.8 & non & \nodata & 2 \\
09293+2143 & NGC2903&  6.2 & 0.83 &  2.3 &$>$0.09& non & \nodata & 2 \\
09320+6134 & UGC05101&160.2& 89.2 & 50.8 & 10.0 & OHM & 1.61 & 8 \\
09517+6954 & M82    &  3.4 &  4.6 &  5.7 & 0.30 & abs+kilo & ($-$1.7) & 9\\
09585+5555 & NGC3079& 16.2 &  4.3 & 24.0 &$\sim$1.0& abs & \nodata & 10\\
10566+2448 & \nodata&173.3 & 93.8 & 61.5 & 10.2 & non & \nodata & 2 \\
11010+4107 & Arp148 &143.3 & 36.5 &$>$47.0& 4.0 & OHM & 1.98 & 11\\
11085+5556 & NGC3556& 10.6 & 1.35 &$>$4.5&$>$0.09& non& \nodata & 2 \\
11176+1315 & NGC3627&  7.6 & 1.26 &  4.4 &$>$0.08& non & \nodata & 12 \\
11176+1351 & NGC3628&  7.6 & 1.01 &  7.1 & 0.24 & abs & \nodata & 4\\
11257+5850 & Arp299 & 43.0 & 62.8 & 29.0 &  2.1 & OHM & 1.38 & 13\\
11460+4859 & NGC3893& 13.9 & 1.15 &  4.1 & 0.23 & non & \nodata & 2 \\
11578$-$0049 & NGC4030& 17.1 & 2.14 & 15.2 & 0.54 & non*& \nodata & 2 \\
11596+6224 & NGC4041& 18.0 & 1.70 &  3.9 & 0.18 & non & \nodata & 2 \\
12239+3129 & NGC4414&  9.3 & 0.81 &  4.6 & 0.16 & non & \nodata & 5 \\
12396+3249 & NGC4631&  8.1 &  2.0 &  2.3 &$\sim$0.08&non&\nodata& 5 \\
12540+5708 & Mrk231 &170.3 &303.5 & 82.2 & 18.6 & OHM & 2.87 & 14\\
12542+2157 & NGC4826&  4.7 & 0.26 &  1.3 & $>$0.04 & non & \nodata & 12 \\
13025$-$4911 & NGC4945&  3.7 &  2.6 &  5.8 &$\sim$0.27& abs & \nodata & 1\\
13086+3719 & NGC5005& 14.0 &  1.4 &  8.2 & 0.41 & non* & \nodata & 2 \\
13135+4217 & NGC5055&  7.3 &  1.1 &  8.6 & $>$0.10 & non & \nodata & 2 \\
13183+3423 & Arp193 & 92.7 & 37.3 & 39.8 &  9.5 & abs & \nodata & 5 \\
13229$-$2934 & NGC5135& 51.7 & 13.8 & 31.3 & 2.73 & non*& \nodata & 7\\
\nodata    & M51    & 9.6  & 4.2  & 19.4 & 0.50 & non & \nodata & 2 \\
13341$-$2936 & M83    &  3.7 &  1.4 &  8.1 & 0.35 & non*& \nodata & 15\\
13428+5608 & Mrk273 &152.2 &129.9 & 65.0 & 15.2 & OHM & 2.53 & 14\\
14306+5808 & NGC5678& 27.8 &  3.0 & 17.2 & 0.75 & non & \nodata & 2 \\
14376$-$0004 & NGC5713& 24.0 &  4.2 &  8.1 & 0.22 & non & \nodata & 2 \\
14514+0344 & NGC5775& 21.3 &  3.8 & 10.9 & 0.57 & abs*& \nodata & 5 \\
15327+2340 & Arp220 & 74.7 &140.2 & 78.5 &  9.2 & OHM & 2.58 & 13\\
16504+0228 & NGC6240& 98.1 & 61.2 & 79.0 & 11.0 & abs & \nodata & 3\\
17208$-$0014 & \nodata&173.1 &234.5 &146.9 & 37.6 & OHM & 3.02 & 16\\
18293$-$3413 & \nodata& 72.1 & 53.7 & 85.5 & 4.03 & non& \nodata & 7\\
18425+6036 & NGC6701& 56.8 & 11.2 & 34.0 & 1.38 & abs & \nodata & 17\\
\nodata    & NGC6921& 60.3 & 11.4 & 17.5 &$\sim$2.81& non& \nodata & 5 \\
20338+5958 & NGC6946&  5.5 &  1.6 &  9.2 & 0.49 & non & \nodata & 2 \\
21453$-$3511 & NGC7130& 65.0 & 21.4 & 44.9 & 3.27 & non& \nodata & 18\\
22132$-$3705 & IC5179 & 46.2 & 14.1 &$\sim$26.4&3.42&non*& \nodata & 7\\
22347+3409 & NGC7331& 15.0 &  3.5 &$>$10.7&$>$0.44& non & \nodata & 2 \\
23007+0836 & NGC7469& 67.5 & 40.7 & 37.1 & 2.19 & abs*& \nodata & 5\tablenotemark{b} \\
23024+1203 & NGC7479& 35.2 &  7.4 & 26.7 & 1.12 & non & \nodata & 5 \\
23365+3604 & \nodata&266.1 &142.0 & 85.0 & 15.0 & OHM & 2.45 & 19\\
23488+1949 & NGC7771& 60.4 & 21.4 & 90.8 &  6.5 & non & \nodata & 5 \\
23488+2018 & Mrk331 & 75.3 & 26.9 & 52.1 & 3.35 & non & \nodata & 2\\
\enddata
\tablecomments{Columns 2--6 are from Table 1 of \citet{gao2004b}.  
References refer to the OH type and luminosity. 
Estimates of the emission line luminosity of OH kilomasers,
corrected for $D_L$ listed in Col.\ 3 and absorption,
are listed in Column 8 in parentheses.}
\tablerefs{
(1) \citet{wg73};
(2) \citet{baan1992}; 
(3) \citet{baan1985a}; 
(4) \citet{rickard82}; 
(5) \citet{schmelz1986}; 
(6) \citet{gallimore1996};
(7) \citet{norris1989}; 
(8) \citet{henkel1990};
(9) \citet{nguyen76};
(10) \citet{haschick1985}; 
(11) \citet{martin1988}; 
(12) \citet{schmelz1988}; 
(13) \citet{baan1989}; 
(14) \citet{baan1985b}; 
(15) \citet{unger1986};
(16) \citet{martin1989}; 
(17) \citet{b1989}; 
(18) \citet{ss92}; 
(19) \citet{bot1990}.}
\tablenotetext{a}{OH types refer to:  
``abs+kilo'' for OH absorption and kilomaser emission; 
``kilo'' for OH kilomaser emission; 
``abs'' for OH absorption; 
``non'' for no OH lines detected; and
``OHM'' for OH megamaser emission.
An asterisk (*) indicates that OH non-detections
have a large rms noise level compared to the typical peak maser line, 
or that the detection of absorption is marginal (3$\sigma$).  These
objects are not plotted in Figures \ref{fig:IR_vs_CO}--\ref{fig:hists}
or used for statistics in the text.}
\tablenotetext{b}{NGC 7469 is listed by \citet{baan1992} as a non-detection, 
but \citet{schmelz1986} claims a 3$\sigma$ detection of OH absorption in a
spectrum with an rms noise 3 times smaller than \citet{baan1992}.}
\end{deluxetable}

\clearpage

\begin{thebibliography}{}
\bibitem[Baan \etal(1985a)]{baan1985a}  Baan, W. A., Haschick, A. D., 
  Buckley, D. \& Schmelz, J. T.  1985, \apj, 293, 394
\bibitem[Baan \etal(1985b)Baan, Haschick, \& Schmelz]{baan1985b}  
  Baan, W. A., Haschick, A. D., \& Schmelz, J. T.  1985, \apj, 298, L51
\bibitem[Baan \etal(1989)Baan, Haschick, \& Henkel]{baan1989}  
  Baan, W. A., Haschick, A. D., \& Henkel, C.  1989, \apj, 346, 680
\bibitem[Baan(1989)]{b1989}  Baan, W. A.  1989, \apj, 338, 804
\bibitem[Baan \etal(1992)Baan, Haschick, \& Henkel]{baan1992}  
  Baan, W. A., Haschick, A., \& Henkel, C.  1992, \aj, 103, 728
\bibitem[Baan, Salzer, \& LeWinter (1998)]{baan1998}
  {Baan, W. A., Salzer, J. J., \& LeWinter, R. D.}  1998, \apj, 509, 633
\bibitem[Baan \& Kl\"{o}ckner(2006)]{baan06}
  {Baan,  W.A. \& Kl\"{o}ckner, H.-R.} 2006, \aap, 449, 559
\bibitem[Bottinelli \etal(1990)]{bot1990}  Bottinelli, L., Gouguenheim, L., 
  Le Squeren, A. M., Martin, J. M., \& Paturel, G.  1990, IAU Circ.\ 4977
\bibitem[Darling \& Giovanelli(2002a)]{darling02a}
  {Darling, J. \& Giovanelli, R.}  2002a, \aj, 124, 100
\bibitem[Darling \& Giovanelli(2006)]{darling06}
  {Darling, J. \& Giovanelli, R.}  2006, \aj, 132, 2596
\bibitem[Gallimore \etal(1996)]{gallimore1996}  Gallimore, J. F., 
  Baum, S. A., O'Dea, C. P., Brinks, E., \& Pedlar, A.  1996, \apj, 462, 740
\bibitem[Gao \& Solomon(2004a)]{gao2004a}  Gao, Y. \& Solomon, P. M.  2004a, 
  \apjs, 152, 63 (GS04a)
\bibitem[Gao \& Solomon(2004b)]{gao2004b}  Gao, Y. \& Solomon, P. M.  2004b, 
  \apj, 606, 271 
\bibitem[Haschick \& Baan(1985)]{haschick1985}  Haschick, A. D. \& Baan, W. A.
  1985, Nature, 314, 144
\bibitem[Henkel \& Wilson(1990)]{henkel1990}  Henkel, C. \& Wilson, T. L.  
  1990, \aap, 229, 431
\bibitem[Krumholz \& Thompson(2007)]{krumholz07}  Krumholz, M. R. \& 
  Thompson, T. A.  2007, \apj, submitted (astro-ph/0704.0792)
\bibitem[Lo(2005)]{lo05} {Lo, K. Y.} 2005, \araa, 43, 625
\bibitem[Lockett \& Elitzur(2007)]{lockett07}
     Lockett, P., \& Elitzur, M. 2007, \apj, submitted
\bibitem[Martin \etal(1988)]{martin1988}  Martin, J. M., 
  Le Squeren, A. M., Bottinelli, L., Gouguenheim, L., \& Dennefeld, M.
  1988, \aap, 201, L13
\bibitem[Martin \etal(1989)]{martin1989}  Martin, J. M., 
  Le Squeren, A. M., Bottinelli, L., Gouguenheim, L., \& Dennefeld, M.
  1989, \aap, 208, 39
\bibitem[Nguyen-Q-Rieu \etal(1976)]{nguyen76}  Nguyen-Q-Rieu, Mebold, U., 
  Winnberg, A., Guibert, J., \& Booth, R.  1976, \aap, 52, 467
\bibitem[Norris \etal(1989)]{norris1989} Norris, R. P., Gardner, F. F., 
  Whiteoak, J. B., Allen, D. A., \& Roche, P. F.  1989, \mnras, 237, 673
\bibitem[Parra \etal(2005)]{parra2005} 
  {Parra, R., Conway, J. E., Elitzur, M., \& Pihlstrom. Y. M.}
  2005, \aap, 443, 383 
\bibitem[Pihlstr\"{o}m \etal\ (2001)]{pihlstrom01} 
  Pihlstr\"{o}m, Y. M., Conway, J. E., Booth, R. S., Diamond, P. J., 
    \& Polatidis, A. G. 2001, \aap, 377, 413 
\bibitem[Rovilos \etal\ (2003)]{rovilos03}  
  Rovilos, E., Diamond, P., Lonsdale, C. J., Lonsdale, C. J., \& Smith, H. E.
  2003, \mnras, 342, 373
\bibitem[Rickard \etal(1982)Rickard, Turner, \& Bania]{rickard82} 
  Rickard, L. J., Turner, B. E., \& Bania, T. M.  1982, \apj, 252, 147
\bibitem[Schmelz \etal(1986)]{schmelz1986}  Schmelz, J. T., Baan, W. A., 
  Haschick, A. D., \& Eder, J.  1986, \aj, 92, 1291
\bibitem[Schmelz \& Baan(1988)]{schmelz1988}  Schmelz, J. T. \& Baan, W. A.
  1988, \aj, 95, 672
\bibitem[Solomon \etal(1997)]{solomon97}  Solomon, P. M., Downes, D., 
  Radford, S. J. E., \& Barrett, J. W.  1997, \apj, 478, 144
\bibitem[Staveley-Smith \etal(1992)]{ss92}  Staveley-Smith, L., Norris, R. P., 
  Chapman, J. M., Allen, D. A., Whiteoak, J. B., \& Roy, A. L.  1992, \mnras, 
  258, 725
\bibitem[Unger \etal(1986)]{unger1986}  Unger, S. W., Chapman, J. M., 
  Cohen, R. J., Hawarden, T. G., \& Mountain, C. M.  1986, \mnras, 220, 1P
\bibitem[Vignali \etal(2005)]{vignali05}
  {Vignali, C. Brandt, W. N., Comastri, A., \& Darling, J.}  
  2005, \mnras, 364, 99
\bibitem[Whiteoak \& Gardner(1973)]{wg73}  Whiteoak, J. B. \& Gardner, F. F.
  1973, \aplett, 15, 211
\end{thebibliography}
\end{document}